\documentclass[conference]{IEEEtran}
\ifCLASSINFOpdf
   \usepackage[pdftex]{graphicx}
\else
   \usepackage[dvips]{graphicx}
\fi

\usepackage[cmex10,fleqn]{amsmath}
\usepackage{units}
\usepackage{hhline}
\usepackage{bbold}
\usepackage{placeins}
\usepackage{bm}
\usepackage[vlined,boxed]{algorithm2e}
\usepackage{kpfonts}
\usepackage{amsmath} % assumes amsmath package installed
\usepackage{amssymb}  % assumes amsmath package installed
\usepackage{kpfonts}

    \newcommand{\cZ} {\mathcal{Z}}

\renewcommand{\bf} {\mathbf{f}}

   \newcommand{\bx} {\mathbf{x}}

  \renewcommand{\phi} {\varphi}

\usepackage{braket}
\usepackage{amsmath}
\IEEEoverridecommandlockouts

\begin{document}

\title{Quantum Prediction of Transport Dynamics in Discretized State Spaces}

\author{\IEEEauthorblockN{Felix Govaers}
\IEEEauthorblockA{Fraunhofer FKIE\\%Dept. Sensor Data and Information Processing\\
Wachtberg, Germany\\
Email: felix.govaers@fkie.fraunhofer.de}
% \thanks{Big thanks to Altamash Khan and Martin Ulmke for the fruitful discussions which lead to the results of this paper.}
}


%

% \noindent\begin{keywords}
% classical mechanics, accumulated state density, smoothing, information functional
% \end{keywords}
% \vspace{1\baselineskip}}

\maketitle

\begin{abstract}
We propose a gate-based quantum algorithm for the prediction step of Bayesian state estimation based on the Fokker--Planck equation on a discretized position--velocity state space. The probability density is encoded in the amplitudes of a quantum state, enabling a compact representation of high-dimensional distributions. Exploiting the circulant structure of finite-difference operators, the evolution is realized in the spectral domain using quantum Fourier transforms and phase rotations.

A key result is that the drift component can be implemented exactly in amplitude space, leading to an accurate reproduction of the classical transport dynamics. In contrast, the diffusion term does not admit a linear representation in amplitude space due to the nonlinear relation between probability density and wave function. To enable a quantum implementation, we introduce a unitary surrogate based on a Wick rotation, transforming diffusion into a dispersive phase evolution. This yields a fully unitary propagation that can be implemented efficiently on a gate-based quantum computer. The proposed method is evaluated numerically for different scenarios and shows strong agreement with the exact solution of the Fokker--Planck equation.
The approach demonstrates the potential of quantum computing for Bayesian state estimation, as the representable state space grows exponentially with the number of qubits. This allows the efficient representation and propagation of probability densities that would otherwise require complex tensor decompositions on classical hardware, making the method a promising candidate for high-dimensional filtering problems.
\end{abstract}

\section{Introduction} % (fold)
\label{sec:introduction}
Bayesian state estimation is a central paradigm for the inference of dynamical systems under uncertainty. In target tracking and related applications, the goal is to estimate the probability density function (pdf) of the system state based on a sequence of measurements. The resulting posterior density $p(\bx_k|\cZ^k)$ represents all available information about the state $\bx_k$ at time $t_k$, conditioned on the measurement history $\cZ^k$. From this density, statistics such as the mean and covariance matrix can be derived.

The recursive Bayesian estimation scheme consists of two fundamental steps: prediction and filtering \cite{b:BarShalom01,b:koch2014,b:govaers}. The prediction step propagates the posterior density forward in time according to the system dynamics, while the filtering step incorporates new sensor information via Bayes' theorem. In this work, we focus exclusively on the prediction step, which is governed by the Chapman--Kolmogorov equation. For continuous-time stochastic processes, this equation admits an equivalent differential formulation in terms of the Fokker--Planck equation.

Classical approaches such as Kalman filtering and its nonlinear extensions, as well as particle filtering methods \cite{b:beyondKF2004, Daum2024}, have been highly successful in a wide range of applications. However, these approaches rely either on restrictive assumptions on the underlying distributions or on sampling-based approximations that can become computationally expensive in high-dimensional settings. An alternative formulation is given by Bayesian state estimation on discretized state spaces, where the probability density is represented explicitly on a grid \cite{j:challa_bar-shalom_fpe, c:fusion16_tensor}. In this setting, the prediction and update steps can be performed directly on the full probability distribution, without relying on parametric assumptions or sampling. While historically limited by computational constraints, this approach has become increasingly attractive due to advances in computational power, parallel processing architectures, and tensor decompositions \cite{j:tensor_CPD_MTT} that allow a compressed representation. 

Despite its conceptual simplicity, the direct numerical solution of the Fokker--Planck equation \cite{Risken1989} on discrete grids remains computationally demanding \cite{Gehlen2025}. At the same time, \emph{quantum computing} offers the ability to represent and transform entire distributions through coherent operations on amplitude-encoded states \cite{Govaers2024}. While others have exploited a quantum wavefunction representation of the conditional densities \cite{Koch2018}, this raises the question of whether the structured nature of grid-based Bayesian estimation can be exploited to directly design efficient quantum algorithms for the prediction step.

A first quantum approach to the Bayesian prediction step was presented in our earlier work \cite{c:govaers_q_pred}. There, the prediction problem was formulated for continuous-noise motion models and addressed by an explicit quantum simulation \cite{Lloyd1996} of Brownian motion and drift. The key idea was to represent the pdf by squared amplitudes of a prepared quantum state and to realize diffusion by repeated applications of a \texttt{PlusMinus} circuit that simulates one step of a discrete random walk. Drift was incorporated by means of controlled repeated \texttt{Plus} and \texttt{Minus} operations on the position register. The previous paper therefore provided an exact quantum-simulation viewpoint for drift--diffusion models, but the number of controlled shifts required scaled linear in the dimension of the velocity state space, that is exponentially in the number of qubits in the velocity register.

The present work differs from that approach in several essential respects.  Instead of simulating state transitions by repeated controlled shift operations, we exploit the circulant structure of the finite-difference derivative matrices and diagonalize the spatial transport operators in the Fourier domain. 

A second important difference concerns the relation between the implemented quantum dynamics and the exact Bayesian prediction operator. In the previous work, the emphasis was on quantum simulation of the underlying stochastic process itself, including diffusion generated by Brownian increments and drift represented through conditioned shift gates. In the present paper, we instead evolve the amplitude representation $\psi=\sqrt{p}$ under a unitary operator derived from the drift generator and then reconstruct an approximate density as $|\psi'|^2$. This yields a quantum algorithm that is significantly more structured and substantially faster. At the same time, the method is no longer exact with respect to the classical prediction density $p'=\exp(\Delta t L)p$, and the analysis of the discrepancy between $p'$ and $|\psi'|^2$ becomes a theoretical contribution of the paper. In this sense, the present work can be understood as a complementary development: while the earlier paper established a direct quantum-simulation framework for stochastic prediction, the current paper introduces a Fourier-based gate construction for deterministic drift that gains efficiency by accepting and characterizing an approximation at the density level. 

Another closely related line of work is the recently proposed quantum algorithm for the diffusion step of grid-based Bayesian filters by Choe et al \cite{Choe2026}. This approach addresses the computational bottleneck of the prediction step from a different perspective, namely by focusing on the convolution structure of the process noise. Their method encodes both the advected state density and the process noise density into separate quantum registers and performs a QFT-based quantum addition. The addition is implemented via a sequence of quantum Fourier transforms and conditional phase rotations, effectively realizing a modular addition in the computational basis. In their work the QFT is used in to implement addition (and thus convolution), whereas in our work it is used to diagonalize the transport operator and enable an efficient spectral propagation. As a consequence, the two methods can be viewed as complementary building blocks for a fully quantum realization of the Bayesian prediction step: the approach of Choe et al. provides an efficient solution for diffusion, while our Fourier-based phase-shift method addresses deterministic drift.

In this work, we address the question of quantum algorithms for density prediction by considering the Fokker--Planck equation. We develop a quantum algorithm that exploits the Fourier diagonalization of the underlying differential operators, enabling the propagation to be implemented as a sequence of quantum Fourier transforms and phase rotations acting on the amplitude representation of the density. While the drift component is realized exactly in amplitude space, the diffusion term is incorporated via a unitary surrogate obtained through a Wick rotation \cite{Peskin1995}, resulting in a dispersive phase evolution that can be efficiently implemented on a gate-based quantum computer.
The central idea of this work is to reinterpret the Fokker–Planck operator as a unitary spectral evolution on amplitudes, enabling the prediction step of Bayesian filtering to be implemented using quantum Fourier transforms and phase rotations.

The contributions of this paper are therefore threefold. First, we derive a gate-based quantum algorithm for the prediction of discretized phase-space densities governed by drift and diffusion dynamics using spectral methods \cite{Trefethen2000}. Second, we provide a theoretical analysis of the approximation error arising from the amplitude representation and the unitary treatment of diffusion. Third, we validate the approach numerically on a quantum simulator and demonstrate that the resulting quantum propagation yields accurate and consistent approximations of the classical Fokker--Planck evolution across different scenarios.

\section{Formulation of the Problem}
\label{sec:formulation_problem}

Bayesian state estimation provides a principled framework for recursively estimating the state of a dynamical system from noisy observations. In contrast to classical approaches such as Kalman filtering or particle filtering, grid-based methods represent the full probability density function (pdf) over a discretized state space. With increasing computational power and advances in parallelization, such approaches have become increasingly attractive, as they avoid restrictive Gaussian assumptions and allow for highly nonlinear and non-Gaussian models.

The prediction step of the Bayesian recursion propagates the current posterior density forward in time according to a stochastic dynamical model. In continuous time, this evolution is governed by the Fokker--Planck equation, which is equivalent to the Chapman--Kolmogorov integral equation. The \emph{Continuous White Noise Acceleration} (CWNA) model —- also known as the \emph{Nearly Constant Velocity} (NCV) model -— is a stochastic process used in target tracking and state estimation to describe an object moving at a roughly steady speed while accounting for small, unpredictable changes in velocity. The density evolution is given by
\begin{align}
\partial_t p(x,v,t)
=
- v \, \partial_x p(x,v,t)
+ \nu_v \, \partial_{vv} p(x,v,t),
\label{eq:fokker_planck}
\end{align}
where the first term represents deterministic transport (drift) and the second term models stochastic diffusion in velocity space. The method presented here also allows a direct generalization to related models such as diffusion in the position, and also multivariate state estimation with position and velocity state spaces in multiple dimensions can be solved directly by considering separated registers for each dimension.

We discretize the state space on uniform grids with $N_x$ points in position and $N_v$ points in velocity. The probability density is then represented as a matrix
\begin{align}
P \in \mathbb{R}^{N_x \times N_v},
\end{align}
with vectorized form $p = \mathrm{vec}(P) \in \mathbb{R}^{N_x N_v}$ using column-major ordering. The differential operators are approximated by finite differences on periodic grids.

Using these definitions, the discrete Fokker--Planck operator takes the form
\begin{align}
L = - \mathrm{diag}(v)\otimes D_x + \nu_v \, D_{vv} \otimes I,
\label{eq:generator}
\end{align}
where $\otimes$ denotes the Kronecker product, $D_x$ is the first-order derivative in position, and $D_{vv}$ is the second-order derivative in velocity. The exact prediction step is then given by
\begin{align}
p' = \exp(\Delta t \, L)\, p.
\label{eq:exact_prediction}
\end{align}

% \paragraph{Amplitude representation.}
In the quantum setting, the density is encoded in the amplitudes of a quantum state via
\begin{align}
\psi = \sqrt{p},
\end{align}
such that measurement probabilities reproduce the classical density. The evolution is then modeled by a unitary operator acting on $\psi$.

For the drift term, the amplitude obeys the same first-order transport equation as the density.
This is due to the fact that for the real part of the wavefunction it holds that 
\begin{align}
	\partial_t p &= \partial_t (\psi^2) = 2\psi \partial_t \psi
	\\
	 \partial_x p &= \partial_x (\psi^2) = 2\psi \partial_x \psi
\end{align}
Consequently, the drift contribution is propagated using the full generator,
\begin{align}
\psi' = \exp\!\left(\Delta t \, L_{\mathrm{drift}}\right)\psi,
\end{align}
with
\begin{align}
L_{\mathrm{drift}} = - \mathrm{diag}(v)\otimes D_x.
\end{align}
This choice ensures that the induced evolution of $|\psi|^2$ reproduces the classical transport behavior.

In contrast, the diffusion operator does not admit an exact linear representation in amplitude space due to the nonlinear relation $p = |\psi|^2$. To obtain a unitary evolution, we replace the real diffusion coefficient by a purely imaginary parameter,
\begin{align}
\nu_v \rightarrow i q,
\end{align}
which transforms the dissipative diffusion operator into a skew-Hermitian generator. The resulting amplitude evolution is given by
\begin{align}
\psi' = \exp\!\left(\Delta t \left[
- \mathrm{diag}(v)\otimes D_x
+ i q \, D_{vv} \otimes I
\right]\right)\psi.
\end{align}

This reformulation yields a fully unitary propagation in amplitude space, which can be implemented efficiently on a quantum computer. The drift term corresponds to a transport operator, while the diffusion term is replaced by a dispersive operator that induces phase mixing in the spectral domain. The reconstructed density $|\psi'|^2$ therefore provides an approximation of the classical Fokker--Planck evolution.

% \paragraph{Unitary reformulation via spectral methods.}

% \paragraph{Objective of this work.}
In this paper, we develop a gate-based quantum algorithm that combines Fourier-domain phase evolution for drift with a unitary spectral approximation of diffusion. The resulting method enables the coherent propagation of a discretized joint density in position and velocity space, while maintaining a compact circuit representation. We analyze the approximation error introduced by the unitary treatment of diffusion and demonstrate that the resulting dynamics provides a close approximation to the classical Fokker--Planck evolution for a wide range of scenarios.

The overall approximation arises from two independent sources:
(i) the nonlinear mapping between density and amplitudes, and
(ii) the unitary surrogate used to approximate diffusion.
The drift component, in contrast, is represented exactly in amplitude space.

\section{Quantum Algorithm for Drift Prediction via Fourier Phase Shifts}
\label{sec:quantum_algorithm}

In this section, we derive a gate-based quantum algorithm for the drift propagation of the amplitude representation.
The key idea is to exploit the diagonalization of the spatial derivative operator $D_x$ by the discrete Fourier transform, which allows the evolution operator to be implemented as phase rotations in Fourier space.

Let $D_x \in \mathbb{R}^{N_x \times N_x}$ denote the finite-difference approximation of the spatial derivative operator on a periodic grid. 
To derive the spectral representation of the discrete derivative operator, we first consider the cyclic shift operators
\begin{align}
S_- &=
\begin{pmatrix}
0 & 1 & 0 & \cdots & 0 \\
0 & 0 & 1 & \cdots & 0 \\
\vdots &  & \ddots & \ddots & \vdots \\
0 & \cdots & 0 & 0 & 1 \\
1 & 0 & \cdots & 0 & 0
\end{pmatrix},
\\
S_+ &=
\begin{pmatrix}
0 & 0 & \cdots & 0 & 1 \\
1 & 0 & \cdots & 0 & 0 \\
0 & 1 & \ddots & \vdots & \vdots \\
\vdots &  & \ddots & 0 & 0 \\
0 & \cdots & 0 & 1 & 0
\end{pmatrix},
\end{align}
which shift a discrete function by one grid point to the right and to the left, respectively. Since both matrices are circulant, they are diagonalized by the discrete Fourier transform matrix $Q_x$ with entries\footnote{Here, we use the notation that has become standard in quantum computing \cite{b:nielson_chuang}, where the phase sign of the forward transformation is positive. Since most quantum computing frameworks and libraries stick to this notation, it makes an implementation from the formulae easier.}
\begin{equation}
(Q_x)_{mn} = \frac{1}{\sqrt{N_x}} \exp\!\left(2\pi i \frac{mn}{N_x}\right),
\qquad m,n = 0,\ldots,N_x-1.
\end{equation}

To see this explicitly, consider the Fourier basis vector
\begin{equation}
\mathbf{q}_k = \frac{1}{\sqrt{N_x}}
\begin{pmatrix}
1 \\ \omega_k \\ \omega_k^2 \\ \vdots \\ \omega_k^{N_x-1}
\end{pmatrix},
\qquad
\omega_k = \exp\!\left(2\pi i \frac{k}{N_x}\right).
\end{equation}
Applying the shift operators yields
\begin{equation}
S_- \mathbf{q}_k = \omega_k \mathbf{q}_k,
\qquad
S_+ \mathbf{q}_k = \omega_k^{-1} \mathbf{q}_k.
\end{equation}
Thus, the eigenvalues of $S_+$ and $S_-$ are
\begin{align}
\lambda_k(S_-) &= \exp\!\left(+2\pi i \frac{k}{N_x}\right),
\\
\lambda_k(S_+) &= \exp\!\left(-2\pi i \frac{k}{N_x}\right),
\qquad
k=0,\ldots,N_x-1.
\end{align}
Equivalently, the diagonalizations can be written as
\begin{equation}
Q_x^\dagger S_- Q_x = \mathrm{diag}\!\left(\exp\!\left(+2\pi i \frac{k}{N_x}\right)\right)_{k=0}^{N_x-1},
\end{equation}
\begin{equation}
Q_x^\dagger S_+ Q_x = \mathrm{diag}\!\left(\exp\!\left(-2\pi i \frac{k}{N_x}\right)\right)_{k=0}^{N_x-1},
\end{equation}
where $Q_x^\dagger$ denotes the conjugate transpose of $Q_x$. The central finite-difference approximation of the spatial derivative on a periodic grid is given by
\begin{equation}
D_x = \frac{S_- - S_+}{2\Delta x}.
\end{equation}
Since $D_x$ is a linear combination of $S_-$ and $S_+$, it is diagonalized by the same Fourier basis. Its eigenvalues follow directly from those of the shift operators:
\begin{align}
\lambda_k(D_x)
&= \frac{1}{2\Delta x}\left(\lambda_k(S_-) - \lambda_k(S_+)\right) \\
&= \frac{1}{2\Delta x}
\left(
\exp\!\left(+2\pi i \frac{k}{N_x}\right)
-
\exp\!\left(-2\pi i \frac{k}{N_x}\right)
\right).
\end{align}
Using the identity
\begin{equation}
e^{i\theta} - e^{-i\theta} = 2i\sin(\theta),
\end{equation}
we obtain
\begin{equation}
\lambda_k(D_x)
=
\frac{i}{\Delta x}
\sin\!\left(\frac{2\pi k}{N_x}\right).
\end{equation}
Hence, the derivative operator admits the spectral representation
\begin{equation}
Q_x^\dagger D_x Q_x
=
\mathrm{diag}\!\left(
\frac{i}{\Delta x}\sin\!\left(\frac{2\pi k}{N_x}\right)
\right)_{k=0}^{N_x-1}.
\end{equation}
This diagonal form is the basis for the efficient implementation of the drift propagator in Fourier space, since the action of the derivative operator reduces to a phase multiplication for each Fourier mode.

Using the diagonalization of $D_x$, the amplitude operator matrix $L_\psi$ can be rewritten as
\begin{align}
L_\psi 
&= -\mathrm{diag}(v) \otimes D_x \\
&= -\mathrm{diag}(v) \otimes Q_x \Lambda_x Q_x^\dagger,
\end{align}
where $\Lambda_x$ is the diagonal matrix with the elements $\lambda_k(D_x)$.
Using properties of the Kronecker product, this can be expressed as
\begin{equation}
L_\psi 
= (I_v \otimes Q_x)\,
\left(-\mathrm{diag}(v) \otimes \Lambda_x\right)\,
(I_v \otimes Q_x^\dagger),
\end{equation}
where $I_v$ denotes the identity matrix in the velocity space.

Consequently, the evolution operator becomes
\begin{equation}
\exp(\Delta t\, L_\psi)
=
(I_v \otimes Q_x)\,
\exp\!\left(-\Delta t\,\mathrm{diag}(v) \otimes \Lambda_x\bigr)\right)\,
(I_v \otimes Q_x^\dagger).
\label{eq:propagator_factorized}
\end{equation}

% %--------------------------------------------------
% \subsection{Diagonal Evolution in Fourier Space}

Since both $\mathrm{diag}(v)$ and $\Lambda_x$ are diagonal matrices, their Kronecker product is also diagonal. Therefore, the exponential in \eqref{eq:propagator_factorized} is diagonal, with entries
\begin{equation}
\exp\!\left(-\Delta t\, v_j \lambda_k \right)
=
\exp\!\left(-i \frac{\Delta t}{\Delta x}\, 
v_j\sin\!\left(\frac{2\pi k}{N_x}\right) \right)
\end{equation}
for each pair $(k,j)$ corresponding to Fourier mode $k$ and velocity grid point $v_j$.

Let $\psi \in \mathbb{C}^{N_x N_v}$ denote the vectorized amplitude representation of the density. The propagation can then be implemented in three steps:
\begin{enumerate}
\item Transform $\psi$ into the Fourier basis in the spatial dimension:
\begin{equation}
\tilde{\psi} = (I_v \otimes Q_x^\dagger)\, \psi.
\end{equation}

\item Apply the diagonal phase evolution:
\begin{equation}
\tilde{\psi}'_{k,j} = 
% \exp\!\left(-\frac{\Delta t}{2}\, v_j \lambda_k \right)\, \tilde{\psi}_{k,j}.
\exp\!\left(-i \frac{\Delta t}{\Delta x}\, 
v_j\sin\!\left(\frac{2\pi k}{N_x}\right) \right)\, \tilde{\psi}_{k,j}.
\end{equation}

\item Transform back to the original basis:
\begin{equation}
\psi' = (I_v \otimes Q_x)\, \tilde{\psi}'.
\end{equation}
\end{enumerate}

%--------------------------------------------------
\subsection{Quantum Circuit Realization}
% \subsection{Implementation of the Fourier-Space Phase Operator by Conditional Rotations}
% \label{sec:phase_operator_rotations}

After the inverse quantum Fourier transform on the position register, the drift prediction operator becomes diagonal in the joint basis of Fourier modes and velocity states. More precisely, for a Fourier mode index $k$ and a velocity grid point $v_j$, the corresponding phase factor is
\begin{equation}
\exp\!\left(i \phi_{k,j}\right),
\end{equation}
with
\begin{equation}
\phi_{k,j}
=
\frac{-v_j \Delta t}{\Delta x}\,
\sin\!\left(\frac{2\pi k}{N_x}\right).
\label{eq:phase_kj}
\end{equation}
Hence, the diagonal unitary in Fourier space is given by
\begin{equation}
U_\phi
=
\sum_{k=0}^{N_x-1}\sum_{j=0}^{N_v-1}
\exp\!\left(i\phi_{k,j}\right)
\ket{k}\!\bra{k}\otimes\ket{j}\!\bra{j}.
\label{eq:Uphi_diag}
\end{equation}

The goal is to express \eqref{eq:Uphi_diag} in a form that can be synthesized from phase rotations and controlled phase rotations on a gate-based quantum computer.

%--------------------------------------------------
% \paragraph{Separation of the phase into Fourier- and velocity-dependent terms.}

For notational convenience, we introduce the Fourier-dependent coefficient
\begin{equation}
\beta_k
:=
-\frac{\Delta t}{\Delta x}\,
\sin\!\left(\frac{2\pi k}{N_x}\right),
\label{eq:beta_k}
\end{equation}
so that the phase can be written as
\begin{equation}
\phi_{k,j} = \beta_k\, v_j.
\label{eq:phi_beta_v}
\end{equation}
In other words, for every Fourier basis state $\ket{k}$, the phase is linear in the velocity value $v_j$.

We assume that the velocity grid is uniformly discretized according to
\begin{equation}
v_j = v_{\min} + \Delta v\, j,
\qquad
j=0,\ldots,N_v-1.
\label{eq:velocity_grid_affine}
\end{equation}
Substituting \eqref{eq:velocity_grid_affine} into \eqref{eq:phi_beta_v} yields
\begin{equation}
\phi_{k,j}
=
\beta_k v_{\min}
+
\beta_k \Delta v\, j.
\label{eq:phi_affine}
\end{equation}
Thus, the phase consists of a term depending only on the Fourier mode and a term proportional to the integer label $j$ of the velocity state.

%--------------------------------------------------
% \paragraph{Binary expansion of the velocity index.}

Let the velocity register consist of $n_v$ qubits such that $N_v = 2^{n_v}$. Then the index $j$ admits the binary expansion
\begin{equation}
j = \sum_{r=0}^{n_v-1} 2^r b_r,
\qquad
b_r \in \{0,1\},
\label{eq:j_bit_expansion}
\end{equation}
where $b_r$ denotes the value of the $r$-th qubit in the velocity register. Inserting \eqref{eq:j_bit_expansion} into \eqref{eq:phi_affine}, we obtain
\begin{equation}
\phi_{k,j}
=
\beta_k v_{\min}
+
\beta_k \Delta v
\sum_{r=0}^{n_v-1} 2^r b_r.
\label{eq:phi_bitwise}
\end{equation}
This decomposition is crucial, since it expresses the total phase as a sum of contributions that are controlled by the individual velocity qubits.

Accordingly, the diagonal unitary \eqref{eq:Uphi_diag} factorizes as
\begin{equation}
U_\phi
=
U_{\min}
\prod_{r=0}^{n_v-1} U_r,
\label{eq:Uphi_factorization}
\end{equation}
where
\begin{equation}
U_{\min}
=
\sum_{k=0}^{N_x-1}
\exp\!\left(i\beta_k v_{\min}\right)
\ket{k}\!\bra{k}
\otimes I_v
\label{eq:Umin}
\end{equation}
is a phase operator acting only on the Fourier register, and
\begin{equation}
U_r
=
\sum_{k=0}^{N_x-1}\sum_{j=0}^{N_v-1}
\exp\!\left(i\beta_k \Delta v\, 2^j b_j\right)
\ket{k}\!\bra{k}\otimes\ket{j}\!\bra{j}
\label{eq:Ur}
\end{equation}
is the phase contribution controlled by the $r$-th velocity qubit. 
%Here, $b_r(j)$ denotes the $r$-th bit of the integer $j$.

%--------------------------------------------------
% \paragraph{Interpretation as conditional phase rotations.}

The operator $U_{\min}$ in \eqref{eq:Umin} applies, for each Fourier mode $\ket{k}$, a phase
\begin{equation}
\theta_k^{(0)} := \beta_k v_{\min}.
\end{equation}
Hence, $U_{\min}$ is a diagonal phase operator on the Fourier register alone.

For each velocity qubit $r$, the operator $U_r$ applies, conditioned on $b_r=1$, the phase
\begin{equation}
\theta_k^{(r)} := \beta_k \Delta v\, 2^r
\label{eq:theta_kr}
\end{equation}
to the Fourier state $\ket{k}$. Therefore, $U_r$ is a Fourier-register phase operator controlled by the $r$-th qubit of the velocity register.

Since all operators in \eqref{eq:Uphi_factorization} are diagonal, they commute and can be implemented in arbitrary order. On a gate-based quantum computer, diagonal unitaries are synthesized from $R_Z$-type phase rotations and controlled phase gates. The present structure is particularly favorable, since the phase angle depends linearly on the binary digits of the velocity register.

\begin{figure*}[!htb]
	\includegraphics[width=2\columnwidth]{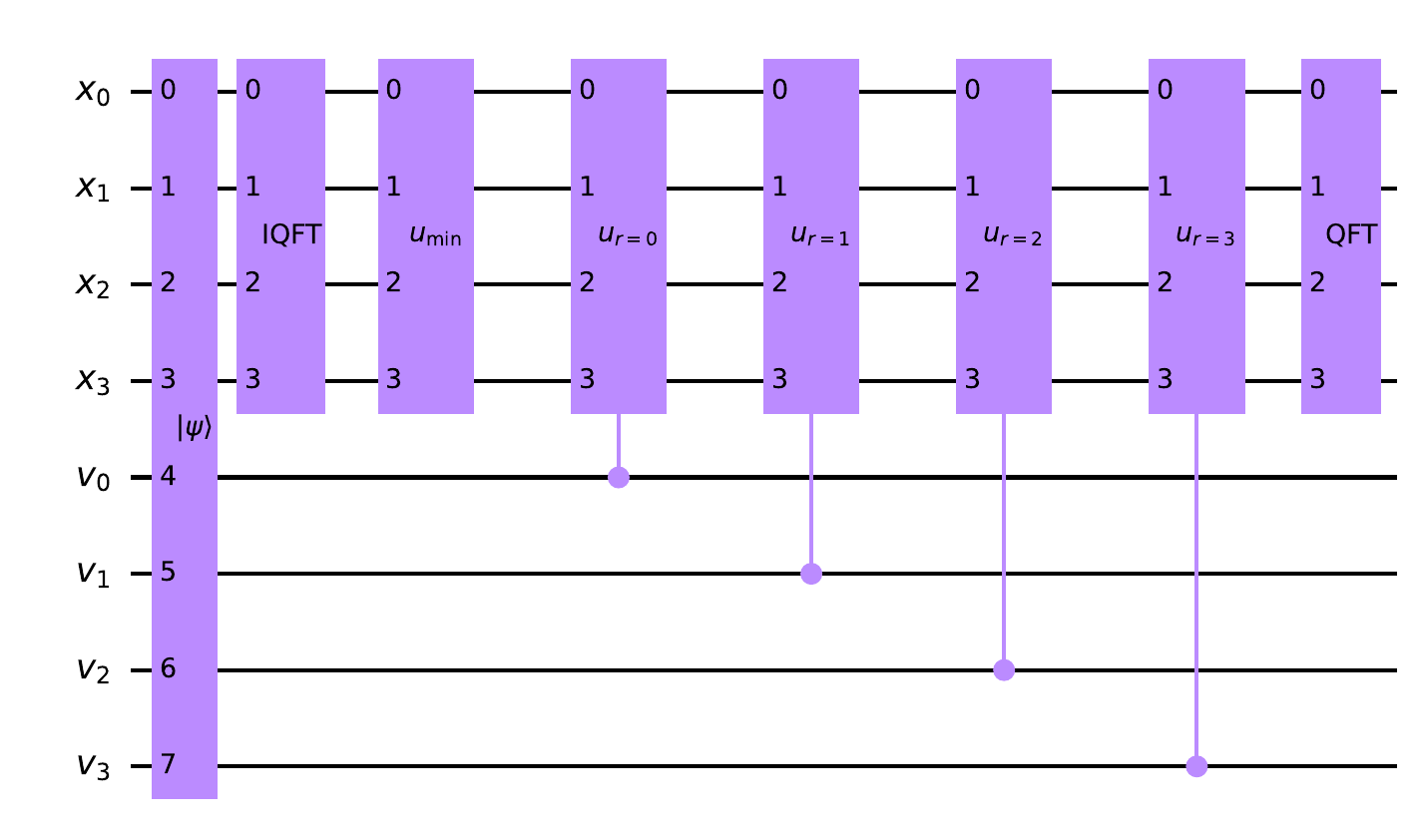}
	\caption{Prediction circuit for $n_x = 4$ qubits for the position and $n_v = 4$ qubits for the velocity register, respectively.}
	\label{fig:circuit}
\end{figure*}

The full drift-prediction circuit is thus composed of the following steps:
\begin{enumerate}
\item Prepare the amplitude-encoded quantum state $\ket{\psi}$ representing the initial density.
\item Apply the inverse quantum Fourier transform on the position register:
\begin{equation}
\ket{\tilde\psi} = (I_v \otimes Q_x^\dagger)\ket{\psi}.
\end{equation}
\item Apply the Fourier-space phase operator
\begin{equation}
U_\phi = U_{\min}\prod_{r=0}^{n_v-1} U_r,
\end{equation}
where $U_{\min}$ acts only on the Fourier register and each $U_r$ is controlled by the $r$-th velocity qubit.
\item Apply the quantum Fourier transform on the position register:
\begin{equation}
\ket{\psi'} = (I_v \otimes Q_x)\, U_\phi \,\ket{\tilde\psi}.
\end{equation}
\end{enumerate}
The resulting circuit for $n_x = 4$ qubits for the position and $n_v = 4$ qubits for the velocity register, respectively, is shown in Figure~\ref{fig:circuit}.

% The remainder of the paper is structured as follows. Section~\ref{sec:drift_model} introduces the discrete formulation of the drift prediction problem. Section~\ref{sec:amplitude_representation} presents the amplitude encoding of the probability density and derives the corresponding unitary amplitude evolution. Section~\ref{sec:quantum_algorithm} describes the quantum circuit implementing the prediction step using a quantum Fourier transform and controlled phase rotations. Section~\ref{sec:error_analysis} analyzes the difference between the exact density evolution and the density reconstructed from the amplitude evolution. Section~\ref{sec:numerical_results} presents numerical experiments that illustrate the behavior of the algorithm and quantify the approximation error. Finally, Section~\ref{sec:conclusion} summarizes the results and discusses possible extensions of the proposed approach.
% \section{Drift Prediction of Discrete Phase-Space Densities} % (fold)
% \label{sec:drift_prediction_of_discrete_phase_space_densities}

% section drift_prediction_of_discrete_phase_space_densities (end)

\section{Quantum Algorithm for Diffusion via Fourier Phase Evolution}
\label{sec:quantum_diffusion}

In this section, we derive a gate-based quantum algorithm for the diffusion component of the prediction step. In contrast to the drift term, which naturally leads to a unitary evolution, the diffusion operator is dissipative and therefore cannot be directly implemented on a gate-based quantum computer. We overcome this limitation by introducing a unitary surrogate based on a spectral transformation of the diffusion operator.

% \subsection{Spectral representation of the diffusion operator}

Let $D_{vv} \in \mathbb{R}^{N_v \times N_v}$ denote the second-order finite-difference approximation of the velocity derivative on a periodic grid\footnote{A diffusion on the $x$ state space is achieved analogously.}. Using the cyclic shift operators $S_+^{(v)}$ and $S_-^{(v)}$, it is given by
\begin{align}
D_{vv} = \frac{S_+^{(v)} + S_-^{(v)} - 2I}{\Delta v^2}.
\end{align}
Since $D_{vv}$ is circulant, it is diagonalized by the discrete Fourier transform matrix $Q_v$,
\begin{align}
Q_v^\dagger D_{vv} Q_v = \Lambda_v
\end{align}
where $\Lambda_v$ is diagonal with its eigenvalues.
% \begin{align}
% \mu_m = -\frac{4}{\Delta v^2} \sin^2\!\left(\frac{\pi m}{N_v}\right),
% \quad m = 0,\ldots,N_v-1.
% \end{align}
%
As above, the eigenvalues of the shift operators are
\begin{equation}
\lambda_m\!\left(S_-^{(v)}\right)
=
\exp\!\left(\frac{2\pi i m}{N_v}\right),
\qquad
\lambda_m\!\left(S_+^{(v)}\right)
=
\exp\!\left(-\frac{2\pi i m}{N_v}\right).
\end{equation}
Since $D_{vv}$ is a linear combination of $S_+^{(v)}$, $S_-^{(v)}$, and the identity, it is diagonalized by the same Fourier basis. Its eigenvalues therefore follow directly:
\begin{align}
\mu_m
&=
\frac{1}{\Delta v^2}
\left(
\lambda_m\!\left(S_+^{(v)}\right)
+
\lambda_m\!\left(S_-^{(v)}\right)
-2
\right)
\\
&=
\frac{1}{\Delta v^2}
\left(
\exp\!\left(-\frac{2\pi i m}{N_v}\right)
+
\exp\!\left(\frac{2\pi i m}{N_v}\right)
-2
\right).
\end{align}
Using the identity
\begin{equation}
e^{i\theta}+e^{-i\theta}=2\cos(\theta),
\end{equation}
we obtain
\begin{equation}
\mu_m
=
\frac{2\cos\!\left(\frac{2\pi m}{N_v}\right)-2}{\Delta v^2}.
\end{equation}
Finally, using\footnote{This can be seen from $e^{i\theta} + e^{-i\theta}=2\cos\theta$ and $e^{i\theta} - e^{-i\theta}=2i\sin\theta$ or from the cosine addition theorem.}
\begin{equation}
2\cos(2\theta)-2 = -4\sin^2(\theta),
\end{equation}
this can be written in the standard form
\begin{equation}
% \boxed{
\mu_m
=
-\frac{4}{\Delta v^2}
\sin^2\!\left(\frac{\pi m}{N_v}\right),
\qquad
m=0,\ldots,N_v-1.
\end{equation}

Thus, the operator $D_{vv}$ admits the spectral decomposition with real, non-positive eigenvalues. 
%This reflects the fact that the second-order derivative acts as a diffusion operator, damping higher Fourier modes more strongly than lower ones.

For a real diffusion coefficient $\nu_v > 0$, the amplitude evolution operator corresponding to diffusion is
\begin{align}
\exp\!\left(\nu_v \Delta t\: D_{vv}\right),
\end{align}
which leads to mode-dependent damping factors in the Fourier domain. Since these factors are real and strictly less than one for $m \neq 0$, the resulting operator is not unitary and cannot be directly realized as a quantum circuit.

% \subsection{Unitary reformulation via Wick rotation}

To obtain a unitary operator, we replace the real diffusion coefficient by a purely imaginary parameter,
\begin{align}
\nu_v \rightarrow i q,
\end{align}
which transforms the diffusion generator into a skew-Hermitian operator. This transformation is closely related to a Wick rotation and yields the unitary evolution
\begin{align}
U_v = \exp\!\left(i \:q \:\Delta t\: D_{vv}\right).
\end{align}

Using the spectral decomposition of $D_{vv}$, the operator can be written as
\begin{align}
U_v = (Q_v \otimes I_x)\,
\exp\!\left(i \:q \:\Delta t\: \Lambda_v \otimes I_x\right)\,
(Q_v^\dagger \otimes I_x).
\end{align}
Since $\Lambda_v$ is diagonal, the exponential is also diagonal, with entries
\begin{align}
\exp\!\left(i q \Delta t \mu_m\right)
=
\exp\!\left(
- i \frac{4 q \Delta t}{\Delta v^2}
\sin^2\!\left(\frac{\pi m}{N_v}\right)
\right).
\end{align}

\subsection{Quantum circuit realization}

The unitary diffusion operator can be implemented using the following three steps:

\begin{enumerate}
\item Apply the inverse quantum Fourier transform on the velocity register:
\begin{align}
|\tilde{\psi}\rangle = (Q_v^\dagger \otimes I_x) |\psi\rangle.
\end{align}

\item Apply a diagonal phase operator in the Fourier basis:
\begin{align}
\tilde{\psi}'_{k,m}
=
\exp\!\left(i q \Delta t \mu_m\right)
\tilde{\psi}_{k,m},
\end{align}
where $m$ denotes the Fourier mode index of the velocity register.

\item Transform back to the original basis:
\begin{align}
|\psi'\rangle = (Q_v \otimes I_x) |\tilde{\psi}'\rangle.
\end{align}
\end{enumerate}

This procedure corresponds to a sequence of quantum gates consisting of an inverse QFT on the velocity register, followed by phase rotations conditioned on the velocity Fourier modes, and a final QFT.

\subsection{Combined drift–diffusion propagation}

Combining the drift operator derived in Section~IV with the unitary diffusion operator yields the full propagation scheme
\begin{align}
|\psi'\rangle = U_x \, U_v \, |\psi\rangle,
\end{align}
where $U_x$ denotes the drift operator acting in the position register and $U_v$ denotes the dispersive diffusion operator acting in the velocity register.

Both operators are diagonal in their respective Fourier domains and therefore admit efficient implementations using quantum Fourier transforms and phase rotations. The resulting circuit consists of two commuting spectral transformations, one in position space and one in velocity space.

% \subsection{Interpretation}

The proposed construction transforms the classical drift–diffusion process into a fully unitary evolution on the amplitude level. While the drift term corresponds directly to the transport operator, the diffusion term is replaced by a dispersive evolution that preserves the norm of the quantum state. This transformation replaces the dissipative smoothing behavior of classical diffusion by phase mixing in Fourier space.

From a physical perspective, the resulting evolution is closely related to Schrödinger-type dynamics, where second-order derivatives give rise to dispersive wave propagation. From a computational perspective, this reformulation enables the efficient implementation of both drift and diffusion using the same circuit primitives, namely quantum Fourier transforms and phase rotations.

The price of this approach is that the reconstructed density $|\psi'|^2$ no longer exactly matches the classical solution of the Fokker--Planck equation. Instead, the method provides a structured approximation whose accuracy depends on the interplay between phase evolution and amplitude reconstruction. The analysis of this approximation error will be addressed in the subsequent section.

\section{Analysis of the Approximation Error}
\label{sec:error_analysis}

In this section, we analyze the discrepancy between the exact classical density evolution governed by the Fokker--Planck equation and the density reconstructed from the quantum amplitude evolution.
The approximation error can be attributed to three distinct effects.

\paragraph{Nonlinearity of the amplitude representation}
The classical evolution acts linearly on $p$, whereas the quantum evolution acts linearly on $\psi = \sqrt{p}$. In general,
\begin{align}
|\exp(\Delta t L_\psi)\sqrt{p}|^2
\;\neq\;
\exp(\Delta t L)\, p.
\end{align}
This mismatch arises because the square operation does not commute with the linear evolution operator. Consequently, interference effects between amplitudes lead to deviations in the reconstructed density.

\paragraph{Unitary approximation of diffusion.}
The classical diffusion operator produces exponential damping of Fourier modes,
\begin{align}
\hat{p}_m' = \exp\!\left(\Delta t \nu_v \mu_m\right)\hat{p}_m,
\end{align}
with $\mu_m \le 0$. In contrast, the quantum algorithm replaces this damping by phase rotations,
\begin{align}
\hat{\psi}_m' = \exp\!\left(i \frac{q \Delta t}{2} \mu_m\right)\hat{\psi}_m.
\end{align}
As a result, diffusion is not realized as smoothing but as dispersive mixing of amplitudes. The damping of high-frequency components is therefore only indirectly reproduced through interference after squaring the amplitudes.

% \subsection{Error representation in Fourier space}

The error can be analyzed conveniently in the spectral domain. Let $\hat{p}_m$ denote the Fourier coefficients of the density and $\hat{\psi}_m$ those of the amplitude. The exact and approximate evolutions are given by
\begin{align}
\hat{p}_m' &= \exp(\Delta t \nu_v \mu_m)\hat{p}_m, \\
\hat{\psi}_m' &= \exp\!\left(i \frac{q \Delta t}{2} \mu_m\right)\hat{\psi}_m.
\end{align}
The reconstructed density involves quadratic combinations of amplitudes,
\begin{align}
\tilde{p}' = |\psi'|^2,
\end{align}
which introduce cross-terms between different Fourier modes. These cross-terms are absent in the classical evolution and represent the primary source of deviation.

For small time steps $\Delta t$, a first-order expansion shows that both evolutions agree up to $\mathcal{O}(\Delta t)$, while differences appear at higher orders. This explains the good approximation observed for moderate time steps.

\paragraph{Error induced by the linear approximation of diffusion}
\label{sec:diffusion_error}

In contrast to the drift term, the diffusion operator does not admit an exact linear representation in amplitude space. This introduces an additional source of approximation error that is intrinsic to the mapping $p = |\psi|^2$.

Consider the diffusion equation
\begin{align}
\partial_t p = \nu_v \, \partial_{vv} p.
\end{align}
Substituting $p = \psi^2$ and assuming $\psi$ to be real-valued for simplicity, we obtain
\begin{align}
\partial_t p = 2\psi\,\partial_t \psi.
\end{align}
On the other hand, applying the second derivative to $p$ yields
\begin{align}
\partial_{vv} p
=
\partial_{vv}(\psi^2)
=
2\psi\,\partial_{vv}\psi
+
2(\partial_v \psi)^2.
\end{align}
Combining these expressions leads to
\begin{align}
\partial_t \psi
=
\nu_v \, \partial_{vv}\psi
+
\nu_v \, \frac{(\partial_v \psi)^2}{\psi}.
\label{eq:psi_diffusion_exact}
\end{align}
The second term on the right-hand side is nonlinear in $\psi$ and cannot be represented by a linear operator.
In the quantum algorithm, the diffusion is approximated by the linear evolution
\begin{align}
\partial_t \psi \approx \nu_v \, \partial_{vv}\psi,
\label{eq:psi_diffusion_linear}
\end{align}
or, in the unitary formulation, by its Wick-rotated counterpart. Comparing \eqref{eq:psi_diffusion_exact} and \eqref{eq:psi_diffusion_linear}, the neglected contribution is
\begin{align}
R(\psi) = \nu_v \, \frac{(\partial_v \psi)^2}{\psi}.
\end{align}
This term represents the intrinsic model error introduced by the linearization.
The resulting error in the density evolution can be characterized by inserting the approximate evolution into $\partial_t p = 2\psi\,\partial_t\psi$. Using \eqref{eq:psi_diffusion_linear}, we obtain
\begin{align}
\partial_t \tilde{p}
=
2\nu_v \psi\,\partial_{vv}\psi,
\end{align}
whereas the exact evolution is
\begin{align}
\partial_t p
=
2\nu_v \psi\,\partial_{vv}\psi
+
2\nu_v (\partial_v \psi)^2.
\end{align}
The discrepancy is therefore given by
\begin{align}
\partial_t p - \partial_t \tilde{p}
=
2\nu_v (\partial_v \psi)^2.
\label{eq:diff_error_density}
\end{align}

This expression provides a direct quantitative interpretation of the error: it is proportional to the squared gradient of the amplitude. Consequently, the error is small in regions where $\psi$ varies slowly and becomes significant in regions with steep gradients.

On a discrete grid, the derivative $\partial_v \psi$ is approximated by finite differences. Denoting the discrete gradient by $D_v \psi$, the error term becomes
\begin{align}
\partial_t p - \partial_t \tilde{p}
\;\approx\;
2\nu_v \, (D_v \psi)^2,
\end{align}
where the square is understood element-wise. This shows that the approximation error scales with the local variation of the amplitude across neighboring grid points.

The analysis reveals that the approximation error introduced by the linear diffusion model is structurally different from the error induced by the unitary reformulation. It is governed by the local gradients of the amplitude and therefore depends strongly on the smoothness of the density. In particular, for Gaussian-like densities with moderate variance, the error remains small, which is consistent with the numerical results presented in this work.

% \subsection{Qualitative behavior}

% Evaluation 
% general parameters
% nqx = 6
% nqv = 6
% delta_x = 1.0
% delta_t = 1
% v_min = -1.5 * 2.5
% v_max = 1.5 *2.5
% nx = 2 ** nqx
% nv = 2 ** nqv
% q = .5
%
% x_grid = np.arange(nx, dtype=float) * delta_x
% mean_x = 0.5 * (x_grid[0] + x_grid[-1])
%
% v_grid = np.linspace(v_min, v_max, nv, dtype=float)
% delta_v = v_grid[1] - v_grid[0]
% mean_v = -1
%
% sigma_x = 0.2 * 0.5 * (x_grid[-1] - x_grid[0])
% sigma_v = 0.1 * 0.5 * (v_grid[-1] - v_grid[0])
%
% scenario 1:
% v=-1
% sigma_v = 0.1 * 0.5 * (v_grid[-1] - v_grid[0])
% results:
% ||P1_from_expm - P_full_diag|| = 0.31982338470314464
% Mean (expm): (np.float64(30.501860135314907), np.float64(-0.9981646377554888))
% Mean (qc):   (np.float64(30.503776178083292), np.float64(-0.9993874960350477))
%
% Covariance (expm):
%  [[40.33312748  0.64349251]
%  [ 0.64349251  0.64351322]]
%
% Covariance (qc):
%  [[40.26110158  0.57324362]
%  [ 0.57324362  0.57506704]]
%
% Mean error: 0.002273016115686192
% Covariance error: 0.14050772897435942
% Total variation: 0.02435907928743335
%
% scenario 2:
% v=2
% results:
% ||P1_from_expm - P_full_diag|| = 0.2800743671839244
% Mean (expm): (np.float64(33.21832113541954), np.float64(1.7183822772254065))
% Mean (qc):   (np.float64(33.36698782081922), np.float64(1.8729306738212839))
%
% Covariance (expm):
%  [[41.50317196  1.81341363]
%  [ 1.81341363  1.81351651]]
%
% Covariance (qc):
%  [[40.65199101  0.96520496]
%  [ 0.96520496  0.96829504]]
%
% Mean error: 0.21444577458668826
% Covariance error: 1.6964151195146044
% Total variation: 0.12049217459833267
\section{Numerical Evaluation}
\label{sec:evaluation}

In this section, we evaluate the proposed quantum algorithm for the prediction step of the Fokker--Planck equation. The numerical experiments are performed on a discretized joint state space in position and velocity. While the algorithm operates on the full joint density $p(x,v)$, we present marginal densities in position and velocity for visualization purposes. All reported metrics, however, are computed on the full joint distribution.

\subsection{Experimental setup}

We consider a discretization with $n_q^x = 6$ qubits for the position register and $n_q^v = 6$ qubits for the velocity register, resulting in $N_x = N_v = 64$ grid points per dimension. The spatial resolution is given by $\Delta x = 1$, and the time step is $\Delta t = 1$. The velocity domain is defined on the interval $[v_{\min}, v_{\max}] = [-3.75, 3.75]$, and the Wick-rotated diffusion parameter is set to $q = 0.5$.

The initial density is chosen as a separable Gaussian distribution
\begin{align}
p(x,v) = p_x(x)\,p_v(v),
\end{align}
with mean position centered in the grid and velocity mean depending on the scenario. The standard deviations are chosen such that the density is smooth and well localized.

% \subsection{Evaluation metrics}

To quantify the approximation quality, we evaluate the following metrics:

\begin{itemize}
\item $L^2$ error between the exact and approximate joint densities,
\item total variation distance,
\item error in the mean,
\item error in the covariance matrix.
\end{itemize}

These metrics provide complementary insights into global accuracy, probabilistic discrepancy, and structural properties of the transported density.

\begin{figure*}[!ht]
	\begin{tabular}{cc}
		\includegraphics[width=0.5\linewidth]{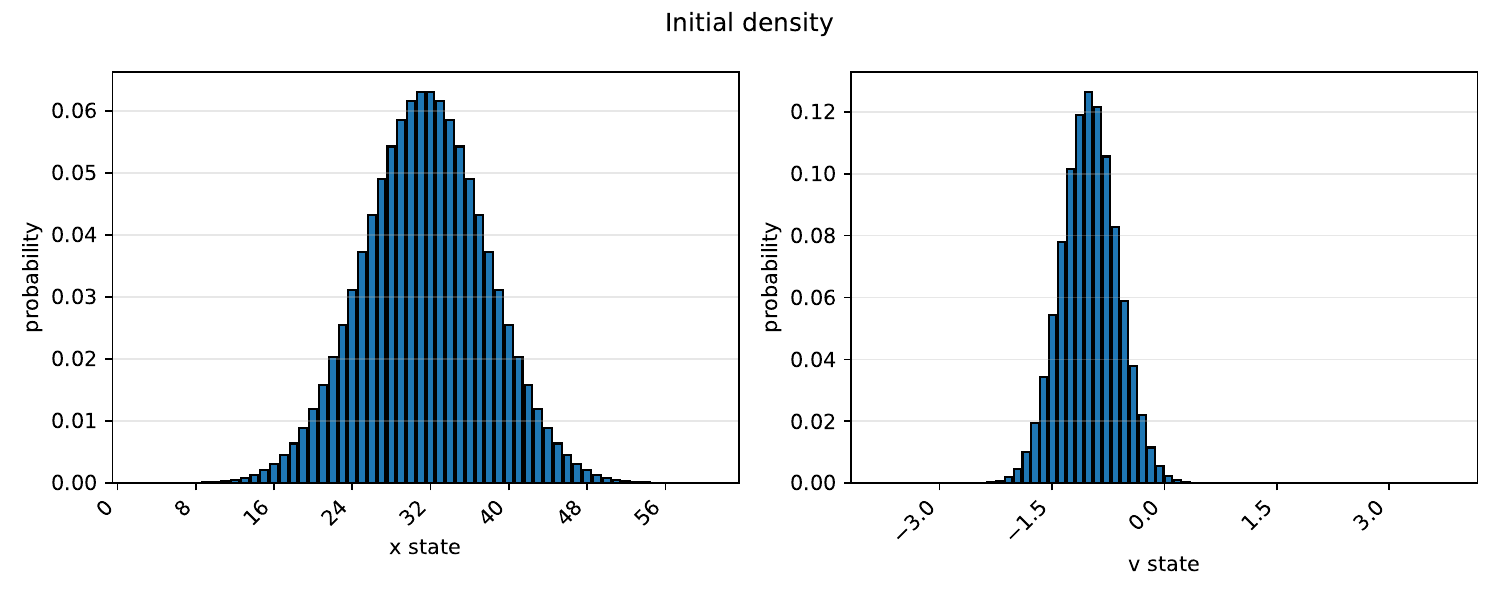}
		&
		\includegraphics[width=0.5\linewidth]{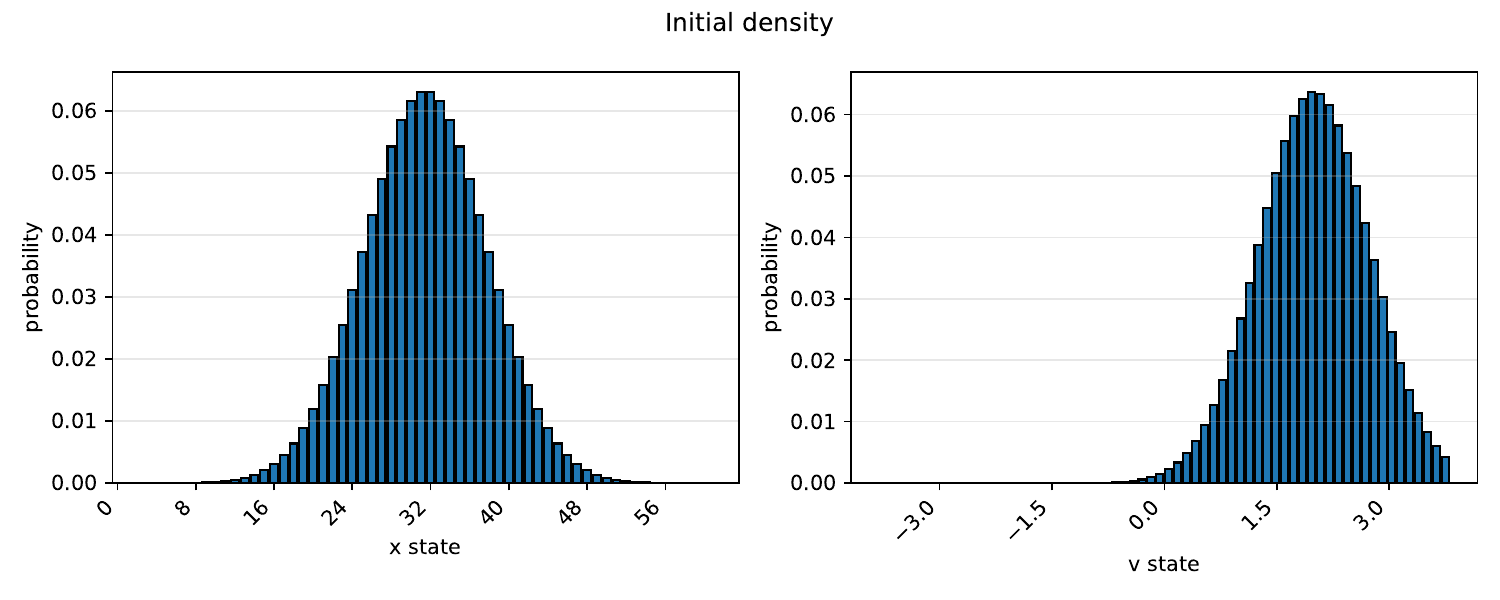}
		\\
		\includegraphics[width=0.5\linewidth]{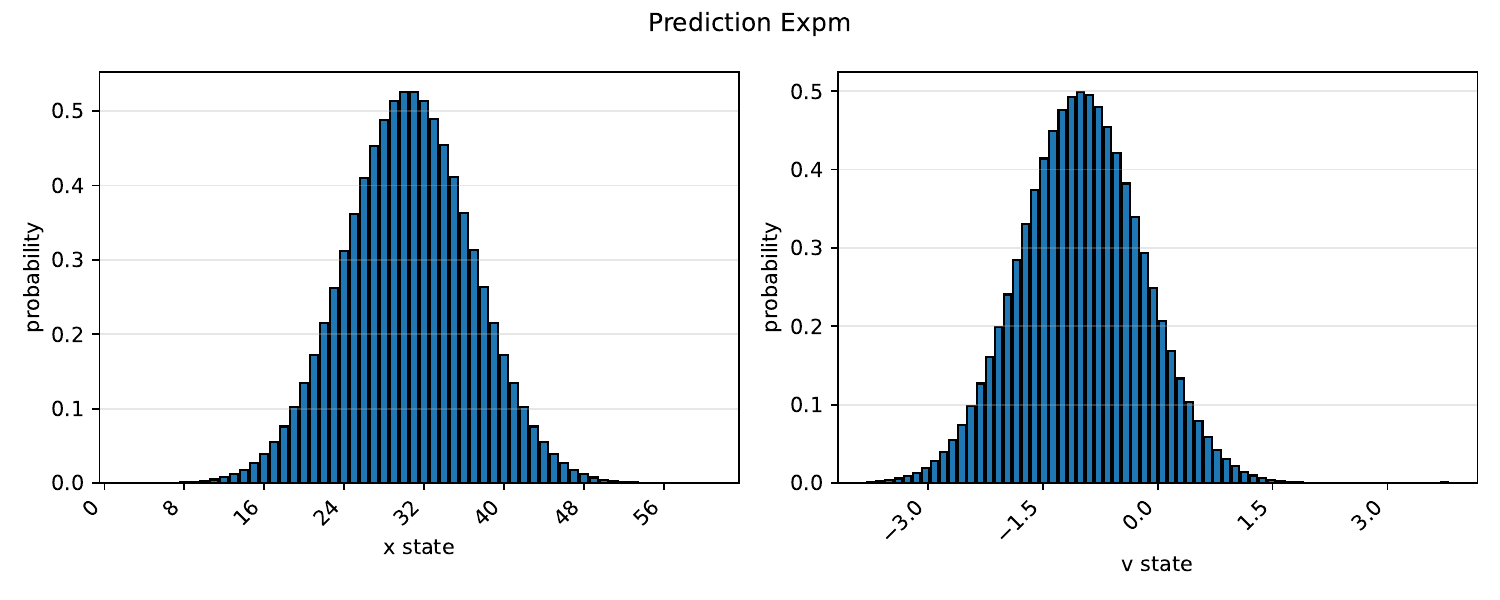}
		&
		\includegraphics[width=0.5\linewidth]{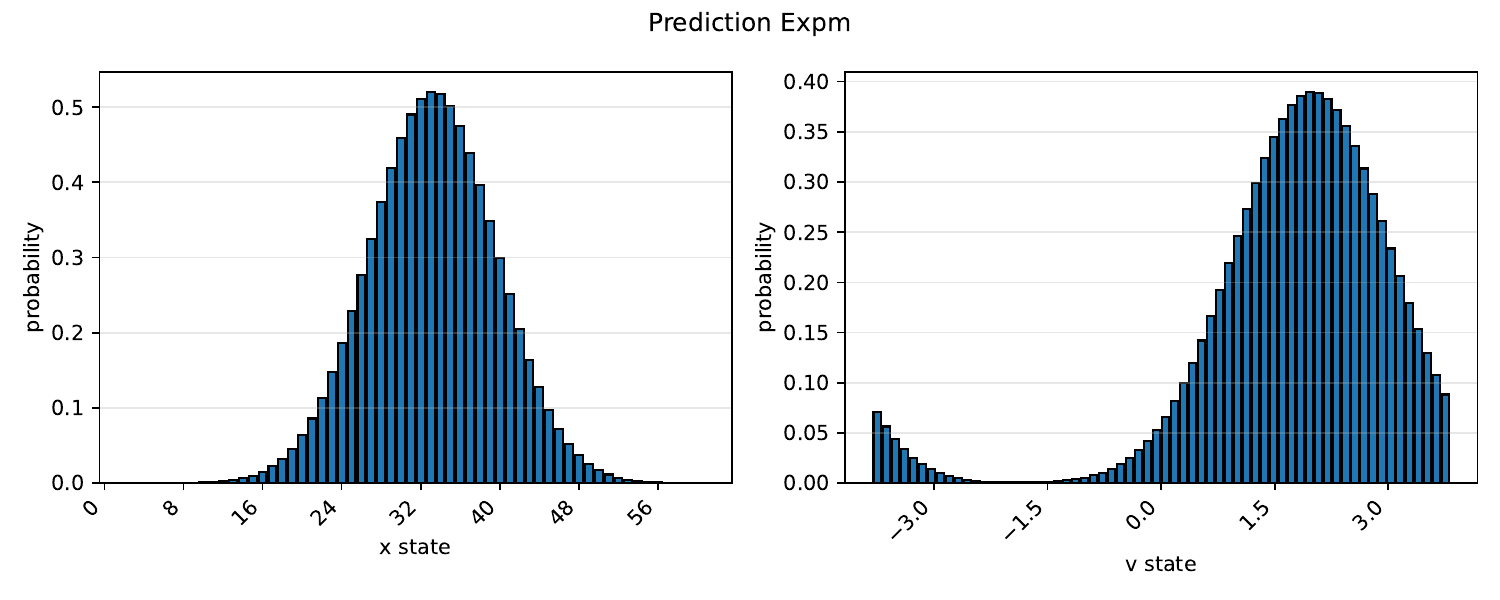}
		\\
		\includegraphics[width=0.5\linewidth]{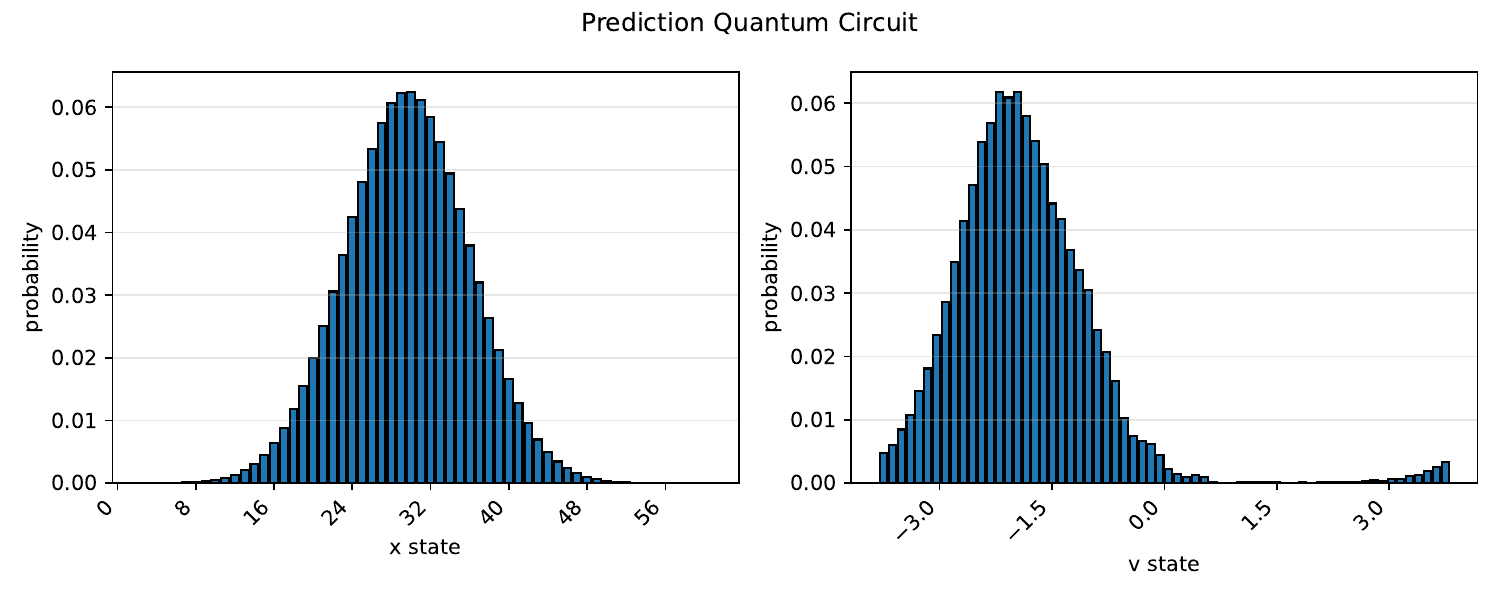}
		&
		\includegraphics[width=0.5\linewidth]{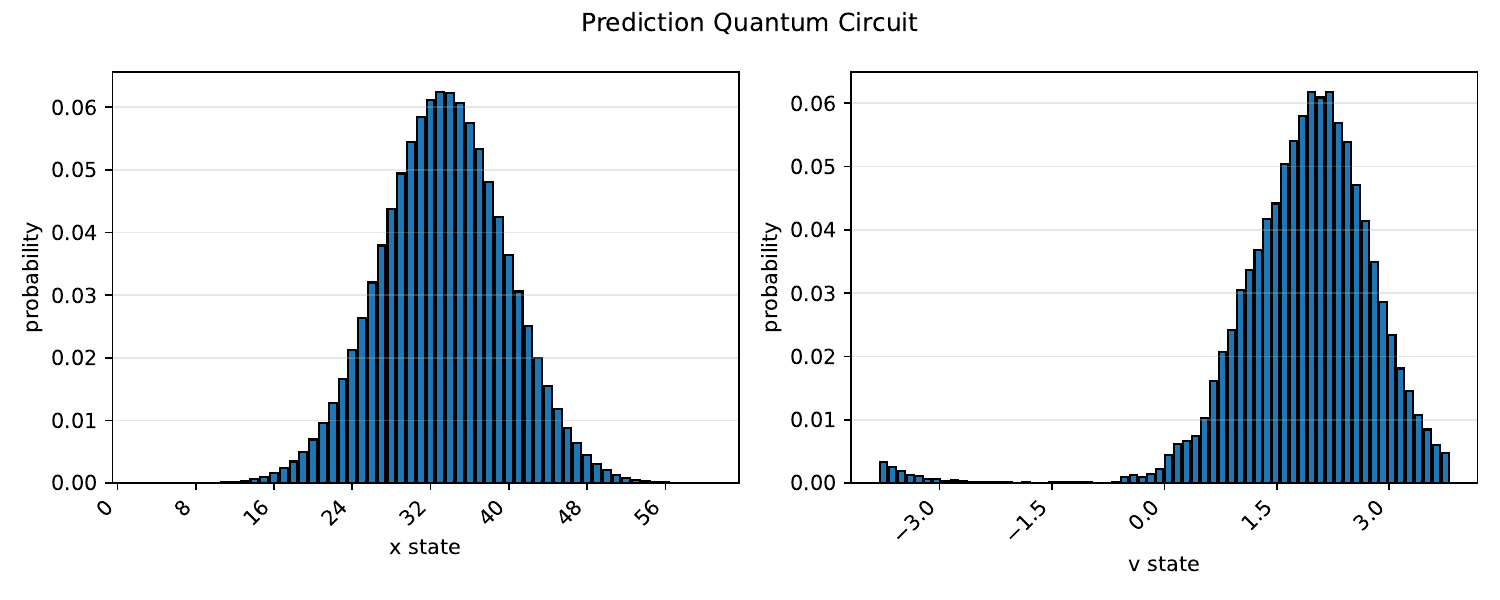}
	\end{tabular}
    \caption{Marginal densities $p_x(x)$ and $p_v(v)$ at the initialization (top), after the exact classical solution using the matrix exponential (middle) and the resulting marginal densities from the quantum circuit (bottom) for Scenario 1 (left) and 2 (right). }
    \label{fig:scenario}
\end{figure*}

In the first scenario, the initial velocity distribution is centered at $v=-1$, corresponding to a moderate drift in negative direction. In the second scenario, the initial velocity distribution is centered at $v=2$, resulting in stronger transport and more pronounced coupling between position and velocity. The corresponding marginal densities are shown in Fig.~\ref{fig:scenario}.

Quantitatively, the results are:
\begin{align*}
\|P_{\text{expm}} - P_{\text{qc}}\|_2 &= 0.3198, \\
\text{Mean error} &= 0.0023, \\
\text{Covariance error} &= 0.1405, \\
\text{Total variation} &= 0.0244.
\end{align*}
for Scenario 1 and 
\begin{align*}
\|P_{\text{expm}} - P_{\text{qc}}\|_2 &= 0.2801, \\
\text{Mean error} &= 0.2144, \\
\text{Covariance error} &= 1.6964, \\
\text{Total variation} &= 0.1205.
\end{align*}
for Scenario 2.

The results show good quantitative and strong qualitative agreement between the quantum approximation and the exact solution. In particular, the mean is reproduced with high precision, indicating that the transport behavior is captured accurately. The covariance error remains small, reflecting a good approximation of the diffusion process. The low total variation distance confirms that the overall density is well approximated.

While the approximation error increases compared to Scenario 1, the qualitative agreement remains good. The marginal densities exhibit consistent transport behavior and spreading, and the main structure of the density is preserved. The increase in error can be attributed to the larger gradients in the density, which amplify the approximation error introduced by the linear diffusion model, as discussed in Section~\ref{sec:diffusion_error}.

Across both scenarios, the proposed quantum algorithm produces results that are highly consistent with the exact solution. The drift component is captured very accurately, as reflected by the small mean errors. The diffusion is approximated well for moderate dynamics and remains qualitatively correct even in more challenging settings.

It is important to emphasize that all computations are performed on the full joint density $p(x,v)$, even though only marginal densities are visualized. The reported metrics therefore reflect the approximation quality in the full state space.

\subsection{Computational Complexity}
\label{subsec:complexity}

In this section, we analyze the computational complexity of the proposed quantum algorithm and compare it to classical grid-based approaches for solving the Fokker--Planck equation.

The discretized state space consists of $N_x$ grid points in position and $N_v$ grid points in velocity. In the quantum setting, the joint density is encoded in a quantum state using $n_x = \log_2 N_x$ qubits for the position register and $n_v = \log_2 N_v$ qubits for the velocity register. Hence, the full state space of size $N_x N_v$ is represented using only $n_x + n_v$ qubits, exhibiting an exponential compression in memory requirements.

The drift propagation is implemented using the following sequence of operations:
\begin{enumerate}
    \item an inverse quantum Fourier transform (QFT) on the position register,
    \item a diagonal phase operator in the joint Fourier--velocity basis,
    \item a forward QFT on the position register.
\end{enumerate}
The QFT on $n_x$ qubits requires $\mathcal{O}(n_x^2)$ elementary gates in its standard decomposition. The diagonal phase operator is realized using controlled phase rotations conditioned on the velocity qubits. Since the phase factor decomposes into contributions from each velocity qubit, this step requires $\mathcal{O}(n_x n_v)$ controlled phase gates. Therefore, the overall complexity of the drift operator is
\begin{align}
\mathcal{O}(n_x^2 + n_x n_v).
\end{align}

The diffusion operator is implemented analogously using a QFT on the velocity register. The required operations are:
\begin{enumerate}
    \item an inverse QFT on the velocity register,
    \item a diagonal phase operator in the velocity Fourier basis,
    \item a forward QFT on the velocity register.
\end{enumerate}
The QFT on $n_v$ qubits requires $\mathcal{O}(n_v^2)$ gates. The diagonal phase operator acts only on the velocity register and requires $\mathcal{O}(n_v)$ phase rotations. Hence, the total complexity of the diffusion operator is
\begin{align}
\mathcal{O}(n_v^2).
\end{align}

Combining drift and diffusion, the total circuit complexity per prediction step is
\begin{align}
\mathcal{O}(n_x^2 + n_v^2 + n_x n_v).
\end{align}
Since $n_x = \log_2 N_x$ and $n_v = \log_2 N_v$, this corresponds to a polylogarithmic scaling in the number of grid points.

\paragraph*{Discussion}
The proposed quantum algorithm therefore achieves an exponential reduction in memory requirements and a polynomial reduction in computational complexity with respect to the number of grid points. This advantage stems from two key properties: the amplitude encoding of the full joint density and the diagonalization of the propagation operators in the Fourier domain.

It should be noted, however, that this complexity advantage applies to the coherent evolution of the quantum state. Extracting the full probability density from the quantum state would require a number of measurements that scales with the size of the state space. In practical applications, one is typically interested in low-order statistics such as means and covariances, which can be estimated efficiently from repeated measurements.

\section{Conclusion}
\label{sec:conclusion}

In this work, we proposed a gate-based quantum algorithm for the prediction step of the Fokker--Planck equation on a discretized position--velocity state space. The method represents the probability density via amplitude encoding and exploits the spectral structure of finite-difference operators to realize the evolution in terms of quantum Fourier transforms and phase rotations.
A key result of this work is that the drift component of the Fokker--Planck equation can be implemented exactly in amplitude space. By propagating the wave function with the full transport generator, the induced evolution of the reconstructed density reproduces the classical advection behavior with high accuracy. This property is confirmed by the numerical results, which show very small errors in the mean across all scenarios.

In contrast, the diffusion component does not admit an exact linear representation in amplitude space due to the nonlinear relation between probability density and wave function. To enable a quantum implementation, we introduced a unitary surrogate based on a Wick rotation of the diffusion operator. This transforms the dissipative diffusion process into a dispersive phase evolution that can be realized efficiently using the same circuit primitives as for the drift. The resulting approximation replaces smoothing by phase mixing in the spectral domain.
The numerical experiments demonstrate that this approach provides a consistent and accurate approximation of the classical prediction step. In particular, for moderate dynamics and smooth densities, the reconstructed probability distributions closely match the exact solution. Even in more challenging scenarios with stronger transport, the method preserves the qualitative structure of the density and yields interpretable deviations that are well explained by the theoretical error analysis.

Moreover, the proposed approach highlights the potential of quantum computing for Bayesian state estimation in high-dimensional settings. Due to the exponential scaling of the representable state space with the number of qubits, even a relatively small quantum system can encode and propagate probability densities that would otherwise require sophisticated tensor decomposition techniques and substantial computational resources on classical architectures.

\bibliographystyle{IEEEtran}
\bibliography{./bibliography.bib}

\end{document}